# A DEEP LEARNING BASED WEARABLE HEALTHCARE IOT DEVICE FOR AI-ENABLED HEARING ASSISTANCE AUTOMATION

FRASER YOUNG[1], LI ZHANG[1], RICHARD JIANG[2], HAN LIU[3] and CONOR WALL[1]

[1]Department of Computer and Information Sciences, Faculty of Engineering and Environment, Northumbria University, Newcastle, UK, NE1 8ST
[2]School of Computing and Communications, Faculty of Science and Technology, Lancaster University, Lancaster, UK
[3]School of Computer Science and Informatics, Cardiff University, Cardiff, UK

**Abstract:**
With the recent booming of artificial intelligence (AI), particularly deep learning techniques, digital healthcare is one of the prevalent areas that could gain benefits from AI-enabled functionality. This research presents a novel AI-enabled Internet of Things (IoT) device operating from the ESP-8266 platform capable of assisting those who suffer from impairment of hearing or deafness to communicate with others in conversations. In the proposed solution, a server application is created that leverages Google's online speech recognition service to convert the received conversations into texts, then deployed to a micro-display attached to the glasses to display the conversation contents to deaf people, to enable and assist conversation as normal with the general population. Furthermore, in order to raise alert of traffic or dangerous scenarios, an 'urban-emergency' classifier is developed using a deep learning model, Inception-v4, with transfer learning to detect/recognize alerting/alarming sounds, such as a horn sound or a fire alarm, with texts generated to alert the prospective user. The training of Inception-v4 was carried out on a consumer desktop PC and then implemented into the AI-based IoT application. The empirical results indicate that the developed prototype system achieves an accuracy rate of 92% for sound recognition and classification with real-time performance.

**Keywords:**
Deep Learning; Convolutional Neural Network; Healthcare; Deafness Assistance; Wearables, Internet of Things

## 1. Introduction

Hearing deficiency is ever prevalent, wherein over one in five human beings are suffering from some degree of hearing loss [1]. This epidemic has been accountable to multiple causes depending on geography. In developing countries including South Africa, South-East Asia and South America, those with the highest prevalence of hearing loss (6.14-7.58%) are frequently due to 'Chronic Ear Infection', over 90% of which occur in these lower economically developed countries [2]. In developed identified countries, hearing loss is often accountable to the likes of old age, and genetic mutations such as the K+-channel gene KCNQ4, having become prevalent in both the Netherlands and the United States [3].

Although hearing-loss does not seem as urgent a research topic as the likes of cancer or blood research at first glance, there are detrimental effects related to deafness affecting all demographics. It has been shown that if congenital hearing loss is not adequately intervened prior to six months from birth, children perform significantly worse in language development tests [4]. Furthermore, it is suggested in the elderly that the prevalence of hearing loss has a direct positive correlation with cognitive functions, suggesting there is a potential link between hearing-loss and onset cognitive decline [5].

Observing the prevalence and effect of hearing loss implores an immediate need for intervention. The purpose of this research is to inform an on-going development of wearable smart-glasses technologies, developed to be affordable and easily deployable to differential subjects, observing current research and applying common and complex methods for audio analysis and classification for real-time visual cues to exchange for auditory stimulation.

Specifically, this research proposes a novel device wherein through utilising open-source micro-controller hardware in combination with 3D printing and deep learning methods for audio classification, results in a low-cost, wearable, smart-glass mountable IoT device that can provide near real-time audio classification of conversation and urban emergency noises such as car horns and emergency sirens.

## 2. Related Work

There are very few developments in wearables tended towards the hard of hearing, and to the best of our knowledge, nothing regarding providing an artificial 'heads-up-display' was capable of real-time translation. However, there have been interventions in those with hearing disabilities, including the adoption of various devices described below.

2.1. Cochlear Implants

Cochlear Implants (CI) are a neural prosthesis conducted wherein an otologist implants a neural interfacing device intent upon modelling the sensory hear cells which transmits vibration signals to the auditory nerve [6]. Despite relative success in the CI, several tribulations have been identified as outlined in [7]. As an example, the contrast between extensive research conducted into the field compared to price is unacceptable, although between 1980-1990 saw an exponential growth in speech recognition. There has been minimal growth shown since (<5%). The high-cost per unit results in particularly low market penetration with only 20% in developed countries and less than 1% in developing countries. Coupled with the highest demography of those with hearing loss in developing regions like Central Asia, South Africa and South America [2], future developments should allow access to the intervention of hearing-loss without aforementioned significant monetary blockades. The device is further illustrated in Figure 1A.

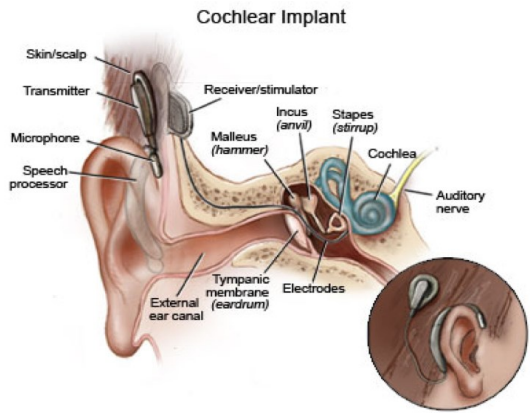

(A)

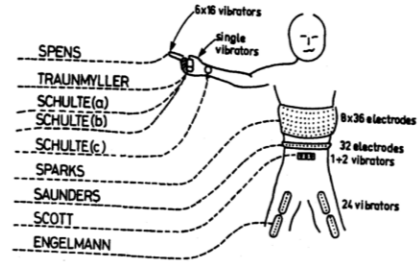

(B)

**FIGURE 1.** Hearing assistant devices: (A) The CI attachment and interaction with the auditory nerve [7]; (B) An illustration of the haptic feedback belt used to represent sound frequencies as described in [8]

2.2. Haptic Feedback Devices

One of the oldest traces of wearables for the deaf results in research conducted in [8] which employed "a wearable tactile sensory aid for profoundly deaf children". Their methodology consists of using a belt worn around the torso attached with Electromagnetic Pulsation Devices (EPD) with the attempt to analyze audios, and output haptic feedback is dependent on the incoming data within the speech frequency spectrum as illustrated in Figure 1B. Their work indicated that after a testing period of four months, speech intelligibility and conversation skills were profoundly improved among children. However, the drawbacks of this methodology were rife including very expensive cost and the professional assistance required in order to properly utilize the device.

Almost 40 years have passed since the above research was conducted and haptic feedback wearables are the most prevalent in representing data through electrical impulses. As early as 2001, research has examined how sensitive humans are by touching a motor-sensory replacement for those with sensory impairment [9]. They developed a non-intrusive wearable device wherein a small hearing-aid-like device is attached to the bone-canal below the earlobe. Audio is taken in and the corresponding electrical signals cause small, specific vibrations to the bone. Their research also found that for those that have suffered from hearing-loss over a large timescale, providing a 'Tactile Display' of information assisted greatly in audio recognition, especially in applications such as lip-reading or conversing [10].

## 3. The Proposed System

Observing existing technologies in related research indicates that there is a significant push towards sensory-replacement. Outside of this field, there is little to nothing to the best of our knowledge. Neither of the observed technologies are particularly accessible to many of those with hearing sensory impairment, especially in CI due to lack of development and high cost. Therefore, a low-cost, non-intrusive system is proposed wherein a prospective user attaches a wearable to a pair of glasses that will take real-time audio signals from a smartphone to interface with Google Cloud servers to display text in observable real-time to the user via the ESP8266 development board over Wi-Fi.

Moreover, the proposed smart-glass device is devised and implemented as an easily deployable, inexpensive device based upon an open-source micro-controller to provide a real-time heads-up display to a prospective user of audio information in the surrounding environment with primary focus on voice synthesis and emergency siren/audio recognition with the intent of warning the user of any danger of the road nearby. The process of development is outlined in Figure 2, demonstrating the full stages of synthesis from 3D model to a workable unit.

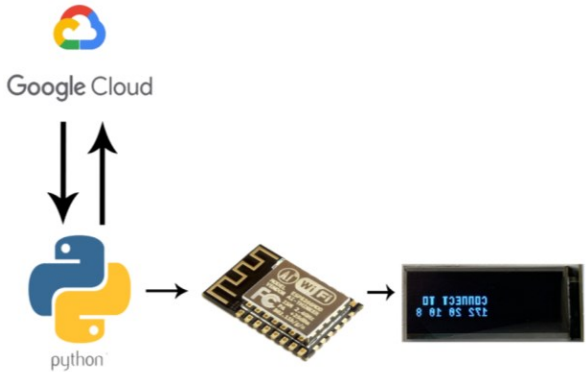

**FIGURE 2.** The design diagram illustrating the communication between all aspects of the product

### 3.1. Sound Classification

The development of the sound classification system to estimate emergency siren weightings in an audio-stream was built utilising methodology and research conducted by Dennis et al. [11], wherein through converting audio signals to Mel-Frequency Cepstral Coefficients (MFCC) displayed through spectrograph representation. Then the spectrograph representation of the incremental audio signals is used as inputs to a deep learning based image classifier for identifying patterns and clusters in represented audios such as siren or fire alarms.

In order to effectively utilise transfer learning to retrain the last few layers of the Inception-v4 deep network, the dataset correlated in 'Dataset for Environmental Sound Classification' [12] is used owing to its extensive library of audio consisting of over 2000 audio recordings of various sound genres. Under the 'Exterior/Urban-Noises' category, exists audio of several useful classifications including sirens, car horns, church bells and chainsaws. The initial experiment is conducted for the detection of emergency sirens such as ambulance and other emergency service alarms, therefore all siren data were put under within an 'emergency' label, while all other data grouped in the 'non-emergency' cluster.

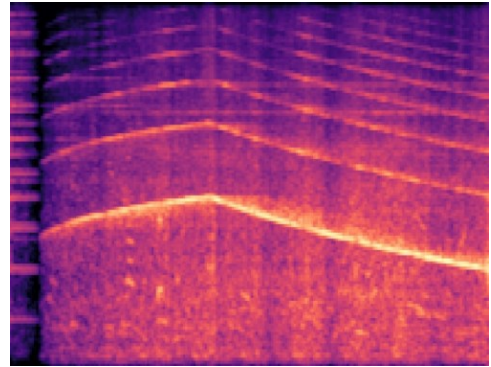

**FIGURE 3.** Example of a positive audio sample containing an emergency tone

To sanitise and process the input dataset, a custom python script was written performing four main tasks, i.e. loading in audio data incrementally from data-containing directories, converting a target audio into the MFCC representation [13-16] as outlined above, converting the frequency data to a spectrograph, and applying post-processing to the graph to ensure all images are of the same resolution and scope of frequency. The images are all saved in their respective labelled directories as JPEG images. An example of data input can be found in Figure 3.

The dataset is split into a typical network training pattern wherein 60% of the images are used for training, 20% for validation and 20% for testing. They are then distributed into TensorFlow's transfer learning platform which interfaces with retraining the final layers of any of Google's developed models.

Inception-v4 [17] was chosen owing to its excellent observed performance with a similar computational cost to its predecessor, Inception-v3. The optimal hyper-parameters are concluded through trial and error with 8000 training steps and a learning rate of 0.01, resulting in an average accuracy rate of 88% in a semi-controlled environment across all classes.

To feed data into the network in near real-time response rate, it was crucial to utilise threading in python. The script consists of two main functions working concurrently split into multiple sub-functions. We establish the thread managing audio input as 'Thread 1'; handling the audio input from the microphone using Python's SoundDevice library in two-second increments. Once it has received two seconds of audio, the data are converted to a spectrograph and processed to be the exact representation of the input audio signal; the image is then saved as a temporary image file. 'Thread 2' handles feeding the information received from Thread 1 into the retrained model. Before the thread is initialised, the TensorFlow model is loaded ahead of time to provide lower latency in real-time use. Using concurrency to feed data into the model improved performance significantly. The last step in the siren classification is to send data to the ESP8266 board.

To achieve this, the ESP board is set up as a slave to receive data from the 'Python Server'. Through using an HTTP request setup, the information is parsed to the ESP board and displayed when a label has been classified. Latency is in general excellent between prediction and display to the user, owing to small data packets; even when an internet connection is slowed, it is received in near real-time.

3.2. Speech Recognition and Classification

The proposed product will only work effectively if the main functionality, speech recognition, is in near real-time performance. To achieve such performance required two iterations of voice processing software, the first utilising Google's offline neural network for speech to text and the second using Google Cloud processing.

The final implementation utilises 'Google Cloud Speech-To-Text' cloud platform, the reason behind which ultimately comes down to speed and can be observed in Figure 4 through mean classification time tests conducted in a controlled environment in order to test classification time of the sentence "The quick brown fox jumps over the lazy dog"; chosen due to its use of every character in the English alphabet.

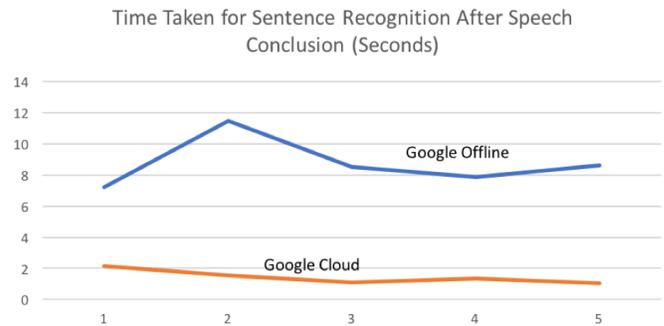

**FIGURE 4.** Tests conducted for quote "The quick brown fox jumps over the lazy dog" across five tests between Google Cloud and Offline processing. As observed, Google Cloud unanimously outperforms the offline processing by up to 9 seconds.

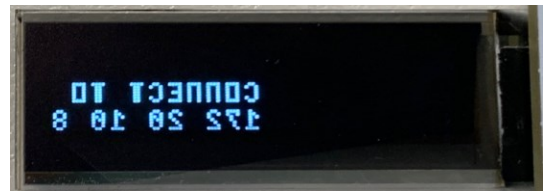

**FIGURE 5.** Example of custom font type displayed on the micro-LCD (Note the text is reversed.)

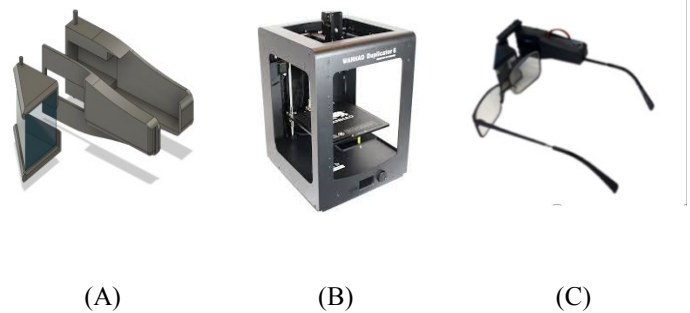

(A)          (B)          (C)

**FIGURE 6.** Conception and production of smart glasses: (A) The custom 3D model developed for this research; (B) The Wanhao Duplicator 6 FDM 3D Printer, used to print the enclosure; and (C) The final product attached to a pair of glasses

The program structure is procedural in nature after defining authentication measures as required by Google Cloud Services. By creating an object to handle microphone input, audio input and prediction are concurrently and procedurally generated. Before the objects are initialised, the user is prompted to input an IP to reach the ESP board; this will open a communication socket between the server and the slave wherein the ESP acts as a slave, hosting an HTML server to receive posted values from the Python server. After the connection is established, the microphone object is initialised and passed into memory as a pointer to audio from the microphone. The validation then takes place to connect to Google's cloud server, concurrently feeding and receiving data to/from the cloud server while taking the microphone input. The server will return a Python array of classifications ordered by prediction weighting, and is displayed through choosing the first index in the array (e.g. predictions[0]) and send it to the ESP board through HTML requests.

### 3.3. Embedded Software

The product is based upon the 'ESP8266 micro-controller', an Internet of Things (IoT) embedded board with the inclusion of 802.11 b/g/n Wi-Fi connectivity and the Serial Peripheral Interface (SPI), useful in attaching serial devices such as LCDs that require a minimum of 3.3v to power. The particular distribution of ESP8266 utilised is the 'HTIT-W8266'; this board was chosen as it has extended features over the base unit including a lithium battery interface and a pre-attached 128x32 LCD. The purpose of the control board in the project is to act as a slave to the Python server, receiving data through an HTML POST request and displaying information to the user.

As the program initialises, the user will be prompted with the IP address of the device on the current WLAN. The server can create a socket to this IP address for hosting information exchange. Once the string data is received via HTML request, the only real function performed is displaying text through the SPI LCD.

In testing, this initially worked to standard, but due to the causality angle of reflection, when a light is passed through a right-angle prism, all text appears backwards in practice.

Examples of the custom texts are illustrated in Figure 5. To account for this, a font-type that can display mirrored text has to be used. But nothing of this type exists. This requires developing a custom font in .XBM format, and a bitmap file with vector scaling.

### 3.4. Component Integration

Figure 2 illustrates the communication and integration between all developed components of the proposed research, and is described below.

As the microphone input is taken from a device, it is sent to both the google cloud voice recognition server or the custom emergency classification server. The servers act concurrently to send and receive data from the corresponding device through decoding and analysing audio to synthesize the desired response. Each time when the new data are received by the device, a buffer update is sent to the ESP8266 via Wi-Fi server, providing near real-time updates of information to the user.

The components are then housed within a 3D printed enclosure as illustrated in Figure 6. The model is designed in AutoCAD Fusion 360 to fit on most glasses frames and to allow for adjustability in conjunction with the end-user. Utilising FDM 3D printing further enabled the low-cost nature of the project and resulted in a quick turn over per prototype; the unit cost including electronic components is £14.92 through Amazon reference prices in 2019.

## 4. Evaluation

As the system has multiple subcomponents, testing and results are split per component. Tests were conducted in a variety of environments as to eliminate any noise bias per location; for example tests were conducted in both a quiet office space to gage a clean signal and also outdoors in a real-world use-case to estimate real-world performance.

### 4.1. Speech Recognition Performance

Due to the significant number of methodologies in synthesizing text from speech, the largest consideration was finding the most appropriate means of doing so. Tests were conducted between Google's offline speech-recognition library included in the SpeechRecognizer Python library and Google's cloud processing (used in Google Assistant products). Figure 4 shows findings of an equivalent comparison of speech-recognition and synthesis time on the string "the quick brown fox jumps over the lazy dog" running on a quad-core mobile i7 with GTX750m. As observed in Figure 7, Google Cloud's processing is unanimously more effective in real-time applications. As illustrated in 'A' (left), the mean-time for classification is almost 8x faster in cloud processing, whereas in

'B' (right), every individual classification rounded at approximately two seconds or less.

### 4.2. Audio Analysis Training Parameters and Performance

The deep learning network for spectrograph image classification was initially trained upon 4,000 training steps with a learning rate of 0.001 as per the recommendation of TensorFlow. The initial training time was 3h 22m running on a quad-core mobile Intel i7 with a GTX760m. We chose the Inception-v4 model due to its more reliable feature extraction and relative ease to train. Final results obtained a controlled-environment accuracy rate of 92%, and a real-world application accuracy rate of 82%, which are discussed comprehensively below.

### 4.3. Discussions

Model performance testing has been performed on two of the most popular image classification models, i.e. Google's Inception-v4 and VGG16 [18]. After training multiple models with different hyper-parameters, Inception-v4 proved to perform the best through testing of the validation dataset, as shown in Table 1. The choice of Inception-v4 is further corroborated through its real-world performance; as seen in Table 2, VGG16's false positive rate (FPR) dramatically increases in comparison with its Inception counterpart, with Inception boosting a higher accuracy and lower FPR in real-world environments; making the choice unanimous as to choose Inception-v4 to train the network.

Hyper-parameters were tuned through a trial-and-error approach. Training steps, learning rate and weight decay were all initialised from random starting variables as per TensorFlow's documentation.

Both models performed the best with the following parameters: Learning Rate = 0.001, Training Steps = 8000, and Weight Decay = 0.05.

It should be seen however, volume advertently affects performance of the model. During validation tests, as the volume increases in the audio segments, accuracy directly correlates alongside lowering of FPR. This is likely due to overfitting of the model through low data differentiation; the majority of emergency noises are a normalised volume through a mix of clean and distorted environments. Due to the normalised volume, the model is likely overfitting loud signals while under-performing in scenarios where the emergency is quiet compared to the surrounding environment. In future work, the dataset must be diversified in order to provide a real-world model that can account for different environments, volumes and tonalities.

Furthermore, hyper-parameter fine-tuning using Particle Swarm Optimisation (PSO) [19-25] could also be explored to further enhance the model performance. Future work will certainly improve upon network optimisation in such a fashion.

### 5. Conclusions

The purpose of this research is to explore how wearables can be applied to help those who are hard of hearing or deaf.

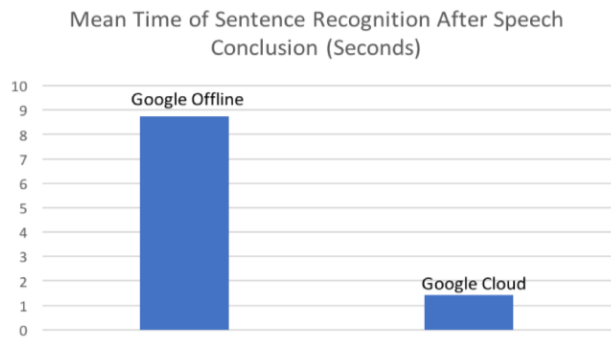

**FIGURE 7.** The average time elapsed for classification between Google's offline and cloud processing packages

**TABLE 1.** Test results of the sound classification for the test dataset when comparing training of Inception-v4 and VGG16, two popular image classification models

|  | **Inception-v4** | **VGG16** |
|---|---|---|
| Accuracy (office validation set) | **92%** | 84% |
| False Positive Rate | **11%** | 13% |

**TABLE 2.** Test results of the sound classification from real-world test dataset as snippets of audio containing emergency noises in a busy environment.

|  | **Inception-v4** | **VGG16** |
|---|---|---|
| Accuracy (real-world validation set) | **82%** | 70% |
| False Positive Rate | **31%** | 52% |

Utilising Google Cloud processing for real-time speech recognition has proved to be extremely effective in the assistive understanding of speech. To validate real-life performance, in future evaluations, it would be beneficial to perform tests with a deaf subject to further understand the experience and take real user feedback.

Although the emergency sound classification shows exceptional performance in controlled environments with synthesized audio, performance is hindered in real-world environments. To expand on this premise further, our work would require a further expansion of dataset with the inclusion of ambient noise and differentiation in data including multiple sirens.

The results of the audio analysis in the real world have equally posed an interesting observation that we aim to look into going forward, as classification accuracy increased in relation to volume of emergency tone; this insinuates one may be able to quantify the 'urgency' of emergency in terms of distance or volume, thus providing a monetary and comparable value as to how significant a danger may be in proximity.

In future work, the proposed system could also be incorporated with more advanced computer vision techniques such as image description generation and visual question answering to provide more detailed description and answer users' enquiries of the abnormal events and emergency scenarios [26-43].